\documentclass{PoS}
\usepackage{slashed}
\usepackage{graphicx}
\usepackage{wrapfig}
\usepackage{color}
\usepackage{amsmath,lineno}
\title{Critical end points in (2+1)-flavor QCD with imaginary chemical potential}

\ShortTitle{QCD phase diagram with imaginary chemical potential}

\author{\speaker{Jishnu Goswami}\\
	Fakult\"at f\"ur Physik, Universit\"at Bielefeld, D-33615 Bielefeld,
Germany;\\
	E-mail:\email{jishnu@physik.uni-bielefeld.de}}

\author{Frithjof Karsch\\
	Fakult\"at f\"ur Physik, Universit\"at Bielefeld, D-33615 Bielefeld,
Germany\\
	Physics Department, Brookhaven National Laboratory, Upton, NY 11973, USA\\
	E-mail: \email{karsch@physik.uni-bielefeld.de}}

\author{Anirban Lahiri\\
	Fakult\"at f\"ur Physik, Universit\"at Bielefeld, D-33615 Bielefeld,
Germany\\
	E-mail: \email{alahiri@physik.uni-bielefeld.de }}
\author{Marius Neumann \\
Fakult\"at f\"ur Physik, Universit\"at Bielefeld, D-33615 Bielefeld,
Germany\\
E-mail: \email{neumann@physik.uni-bielefeld.de  }}

\author{Christian Schmidt\\
	Fakult\"at f\"ur Physik, Universit\"at Bielefeld, D-33615 Bielefeld,
Germany\\
	E-mail: \email{schmidt@physik.uni-bielefeld.de }}

\abstract{We present here the results from an ongoing determination of the critical quark mass
	in simulations of (2+1)-flavor QCD with an imaginary chemical potential. Studies with unimproved actions found the existence of a critical quark mass value at which the crossover transition ends on a second order phase transition and
	becomes first order for smaller values of the quark mass for the case of both vanishing and imaginary chemical potential.
	We use the Highly Improved Staggered Quark (HISQ) action and perform
	calculations in the Roberge-Weiss (RW) plane, where the value of the critical
	mass is expected to be largest. The lowest quark mass value used in our simulation corresponds to the pion mass $m_\pi$,
	down to $40$ MeV. Contrary to
	calculations performed with unimproved actions we find no evidence
	for the occurrence of first order transitions at the smallest quark mass
	values explored so far. Moreover we also show that the chiral observables are sensitive to the RW transition. Our results also indicate that the RW transition and chiral transition could coincide in the chiral limit.
}

\FullConference{Corfu Summer Institute 2018 "School and Workshops on Elementary Particle Physics and Gravity"\\
	(CORFU2018)\\
	31 August - 28 September, 2018\\
	Corfu, Greece}

\begin{document}
\section{Introduction}
The QCD phase diagram at non-zero temperature and baryon chemical potential
plays a central role for our understanding and analysis of the properties
of strongly interacting matter. Large experimental programs at the
Relativistic Heavy Ion Collider (RHIC) in the USA and the Large Hadron
Collider (LHC) in Geneva, Switzerland, are devoted to the study of this 
matter which does undergo
a transition from ordinary hadronic matter at low temperatures and densities
to a quark-gluon plasma at high temperatures and/or densities.

At physical values of the light and strange quark masses it is now well
established that this transition is not a true phase transition but a 
crossover between two physically distinct regimes. On the other hand,
chiral symmetry, which is an exact symmetry in the limit of vanishing
values of the quark masses, is known to be spontaneously broken at
zero temperature and being restored in the perturbative high temperature
limit of QCD. A phase transition thus is expected to occur in the chiral
limit, {\it i.e.} in the limit of vanishing up and down quark masses, at 
some critical temperature $T_c$.

Using lattice QCD (LQCD) calculations it nowadays is possible to perform 
non-perturbative calculations with quark masses much smaller than their 
physical values. A recent LQCD study of $(2+1)$-flavor QCD, {\it i.e.} with 
two light flavors and a strange quark mass fixed to its physical value, 
performed by the HotQCD Collaboration \cite{Ding:2018auz,Ding:2019prx}, has 
determined the chiral transition temperature ($T_c \sim 132$ MeV) in the chiral 
limit for the case of vanishing quark chemical potentials ($\mu$).  It also 
has been shown that the chiral transition is consistent with being a $2^{nd}$ 
order transition that is likely to belong to the $3$-$d$~$O(4)$ universality 
class. 
In more concrete terms no signal for a first order transition has been
found down to a pion mass of about $55$~MeV.
 
For two massless quark flavors the QCD Lagrangian  has an exact 
$SU(2)_L\times SU(2)_R \times U(1)_A \times U(1)_V$ symmetry. 
However at low temperature, the QCD vacuum is not invariant under the flavor 
chiral symmetry. As a consequence $SU(2)_L\times SU(2)_R$ reduces to $SU(2)_V$ 
due to spontaneous symmetry breaking giving rise to three massless Goldstone 
bosons (pions) in the spectrum. The flavor chiral symmetry gets restored at 
high temperature ($T_c$) via a true phase transition in the chiral limit. 
In-spite of the fact that the $U(1)_A$ symmetry is broken by quantum 
corrections at all temperatures, which causes the mass splitting between the 
$\eta$ and $\eta^{\prime}$ mesons at low temperature, at very high temperature 
the quark matter will behave like a free Fermi gas and quantum effects will be 
negligible; the $U(1)_A$ thus will be effectively restored. 
A crucial question is whether this effective restoration already happens at 
temperatures close to $T_c$ or not. In case, $U(1)_A$ is effectively
restored already at $T_c$, the spontaneous breaking of a larger symmetry group,
$U(2)_L\times U(2)_R/U(2)_V$, would be of relevance for the 
chiral phase transition, which according to the standard scenario put
forward by Pisarski and Wilczek \cite{Pisarski:1983ms} may give rise to a 
first order phase transition, although it has been pointed out that a 
second order transition also is possible for this larger symmetry 
group \cite{Pelissetto:2013hqa,Eser:2015pka}. 

\begin{figure}
	\centering
	\includegraphics[page=2,scale=0.22]{phase_diagram_col.pdf}\hspace*{-0.5cm}
	\includegraphics[page=1,scale=0.22]{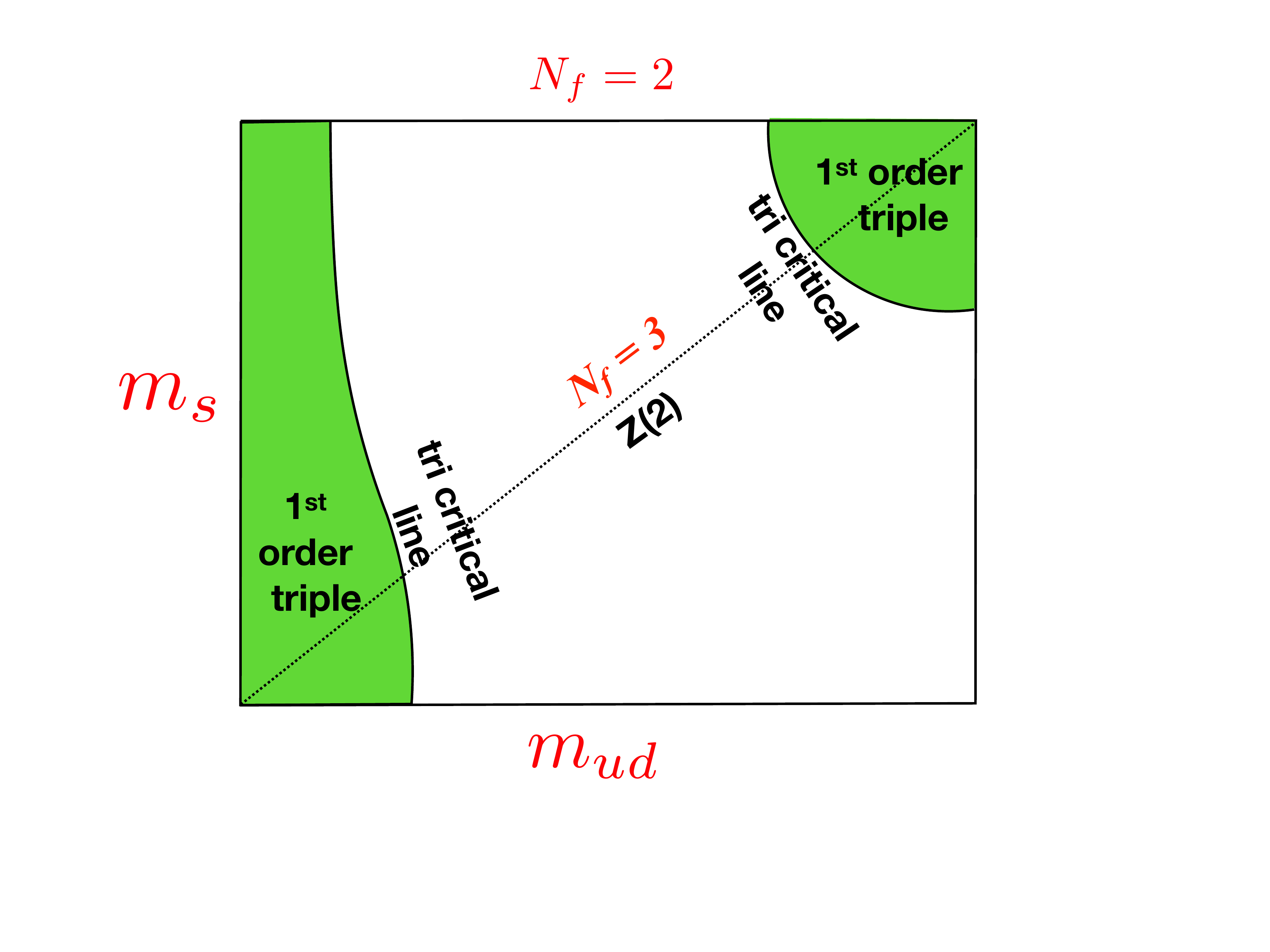}\vspace*{-0.8cm}
	\caption{Sketch of possible phases in the light and strange quark mass 
		plane (Columbia plots) for $\mu=0$ (left) and 
$i\mu/T=i\pi/3$ (right) planes.}
\label{Columbia}
\end{figure}

Studies with unimproved staggered 
fermions~\cite{deForcrand:2010he,Philipsen:2016hkv,Cuteri:2015qkq}, performed
on coarse lattices, suggest that the chiral transition can become $1^{st}$ 
order for reasons presumably totally unrelated to the restoration of
the anomalous $U_A(1)$ symmetry. Below 
a certain critical quark mass first order transitions have been 
found, which led to a possible version of the phase diagram in the light and 
strange quark mass plane (Columbia plot) as shown in Fig.~\ref{Columbia}~(left).
However, in calculations with improved staggered fermion actions, e.g. the
Highly Improved Staggered Quark (HISQ) or 
stout actions, which strongly reduce discretization errors, 
no $1^{st}$ order transition is found in (2+1) \cite{Bonati:2018fvg} as well as 
3-flavor LQCD  \cite{Bazavov:2017xul} even for quite small 
quark masses, corresponding to Goldstone pion masses as small as 
50 MeV. 

If a first order transition exists at smaller quark masses than those
currently explored in calculations at $\mu=0$, such a transition 
would more easily be detected in calculations with a non-vanishing
imaginary chemical potential.
It has been found in calculations with unimproved staggered fermions that the 
range of quark masses, for which a $1^{st}$ order transitions occurs, 
increases when one performs calculations with a non-zero imaginary chemical 
potential (see Fig.~\ref{Columbia}~(right)), i.e. for negative $\mu^2$, 
This $1^{st}$ order region becomes maximal for 
$|i \mu/T|=\pi/3$~\cite{deForcrand:2010he,Philipsen:2016hkv,Cuteri:2015qkq}. 
In this work we report on our ongoing studies of the  QCD phase diagram with 
imaginary chemical potential and small values of the light quark masses, 
keeping the strange quark mass at its physical value. We use the HISQ action 
as our lattice discretized Dirac operator. 
\vspace{-0.3cm}


\vspace{-0.2cm}
\section{Imaginary chemical potential formalism and  Roberge-Weiss (RW) plane}
The QCD partition function with identical imaginary chemical potential, 
$i\hat{\mu}$, for $f$ quark flavors with mass $m_f$
can be written as,
\begin{eqnarray}
	\hspace*{-0.2cm}Z(T,V, \mu)=\int \prod_f D\psi_f D\bar{\psi}_f DA_{\mu} \exp\Big[-\int_{0}^{\beta}\hspace*{-0.2cm} dx_0\int_V d^3x\sum_f {\bar{\psi}_f(\gamma_{\mu}D_\mu+m_f -
	 i\mu\gamma_0)\psi_f-
 \frac{1}{4}F^2_{\mu\nu}  }\Big] .\ \label{eq:img_path_integral1}
\end{eqnarray}
In the Euclidean time direction one imposes anti-periodic boundary condition 
for the quark fields ($\psi(0,{\bf x})=-\psi(\beta,{\bf x})$) and periodic boundary 
condition for the gauge fields $A_{\mu}$. Roberge and Weiss
\cite{Roberge:1986mm} observed that in Eq.(\ref{eq:img_path_integral1}) 
one can substitute 
$\psi_f(x_0,{\bf x})=exp(ix_0\hat{\mu}/\beta)\psi_f(x_0,{\bf x})$, with  
$\hat{\mu}=\mu\beta$, which merely results in a change of boundary conditions
for the fermion fields,
\begin{eqnarray}
	Z(T,V, \mu)=\int \prod_f D\psi_f D\bar{\psi}_f DA_{\mu} \exp\Big[-\int_{0}^{\beta} dx_0\int_V d^3x \sum_f {\bar{\psi}}_f(\gamma_{\mu}D_\mu+m_f) -\frac{1}{4}F^2_{\mu\nu}\Big]\; ,
	\label{eq:img_path_integral2}
\end{eqnarray}
with $\psi(x_0,{\bf x})=\exp(i(\pi+\hat{\mu}))\psi(x_0+\beta,{\bf x})$. 
They also pointed out that in QCD the partition 
function is periodic, $Z(\hat{\mu})=Z(\hat{\mu}+2\pi k/3)$, which arises 
from the global $Z(3)$ invariance inherent in
Eq.(\ref{eq:img_path_integral1}) and Eq.(\ref{eq:img_path_integral2}). 
Roberge and Weiss have argued that in a given $Z(3)$ sector,
$2\pi k/3 \le \hat{\mu} \le 2\pi (k+1)/3$, the partition function
is symmetric around the center of this interval, $\mu/T=(2k+1) \pi/3$;
{\it i.e.} an additional $Z(2)$ arises, which may become spontaneously 
broken at high temperature. The first derivative of free energy 
$F(T,\hat{\mu},V)=-T\ln Z(T,\mu,V)$ may become discontinuous at 
$\mu/T=(2k+1) \pi/3$ and a 
$1^{st}$ order phase transition may occur for these values of the imaginary 
chemical potential at high temperature. These particular choices of the 
imaginary chemical potential in a given $Z(3)$ sector are known as the 
Roberge-Weiss (RW) planes. 

In LQCD formalism one introduces the imaginary chemical potential ($i \mu$) 
by multiplying all time-like link fields with a phase,
$U_{x,\hat{0}}\rightarrow \exp(ia\mu)U_{x,\hat{0}}
~,U_{x,\hat{0}}^{\dagger}\rightarrow \exp(-ia\mu)U_{x,\hat{0}}$ \cite{Hasenfratz:1983ba}. 
The partition function then still has a real and positive fermion determinant. 
The gauge fields variables, $U_{x,\hat{0}}$, can always be transformed by
globally multiplying all time-like gauge field variables with an element of 
the center group,
$z\in\{1,\exp[\pm i\frac{2\pi}{3}]\}$, i.e.
$ U_{x,\hat{0}} \rightarrow z~U_{x,\hat{0}}$, where $z\in\{1,\exp[\pm i\frac{2\pi}{3}]\}$. Under this transformation the partition function is again invariant,
$Z(T,\hat{\mu},V)=Z(T,\hat{\mu}+2\pi/3,V)$. 
\begin{figure}
	\centering
	\includegraphics[scale=0.22]{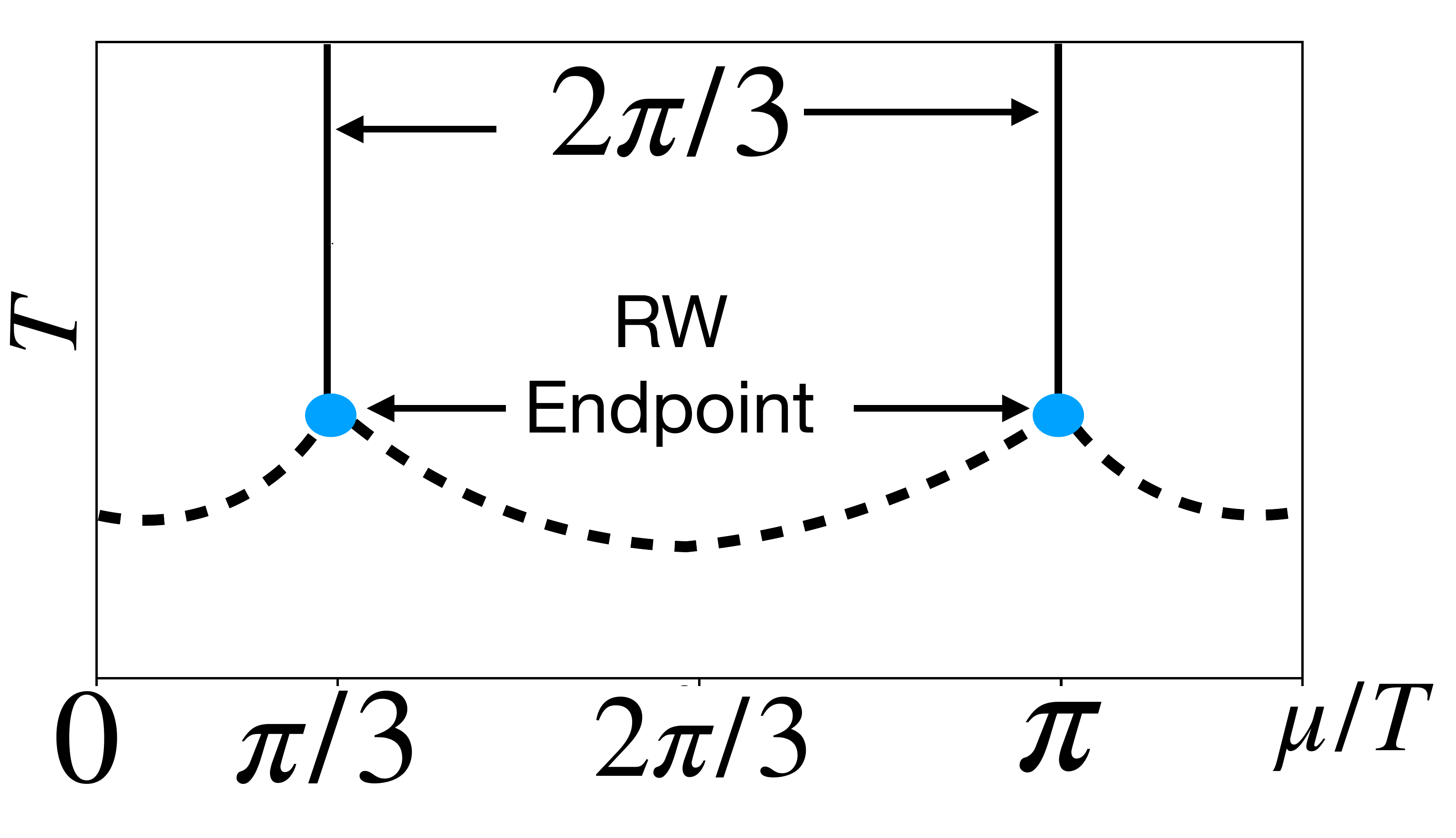}
	\vspace*{-0.2cm}
	\caption{
		A possible 
		phase diagram in the imaginary chemical potential
		plane. Solid lines show first order phase transitions, dotted 
		lines correspond
		to crossover transitions and the blue points indicate $2^{nd}$ 
		order phase transition points.}
	\label{phase_imL}
\end{figure}
\noindent

As an order parameter for the spontaneous
breaking of the $Z(2)$ symmetry in the RW plane the expectation
value of the imaginary part of the Polyakov loop is commonly used,
\begin{equation}
        \langle {\rm Im}~ P\rangle = \lim_{\mu\rightarrow\mu_{RW} ,
	{V\rightarrow \infty}} \left\langle \frac{1}{N_\sigma^3} 
	\sum_{\bf x}{\rm Im}
	{\rm Tr}\prod_{x_0=1}^{N_\tau} U_{x,\hat{0}}\right\rangle
        \, .
	\label{Polyakov}
\end{equation}
A specific scenario for phase transitions in the RW plane, consistent with the results presented here, is shown in 
Fig.~\ref{phase_imL}. 
Alternative scenarios have been found in 
calculations with standard staggered and also Wilson fermions,
where the RW endpoint turns out to be a triple point for sufficiently
small values of the light quark masses and three first order transition lines
would emerge from this triple point.
For $N_f =2$ a $1^{st}$ order triple point 
has been found  on coarse lattices with temporal
extent $N_\tau =4$ in calculations with standard staggered fermions below 
$m_\pi^{cri}\sim 400$ MeV  and with the standard Wilson fermion action below
$m_\pi^{cri}\sim 930$ MeV
\cite{deForcrand:2010he,Philipsen:2016hkv,Cuteri:2015qkq}.
Similar to the case of vanishing chemical potential, results turn out to be 
quite different in calculations performed 
with improved actions. No $1^{st}$ order transition has been found at least 
for $m_{\pi} >50$ MeV in calculations with the stout improved 
action~\cite{Bonati:2018fvg,Bonati:2016pwz}. Some calculations have also been
performed using the 2-flavor HISQ action. In this case no clear-cut results on 
the order of the phase transition have been reported so far \cite{Wu:2018oed}.
In our calculations we use the $(2+1)$-flavor HISQ action together with an
${\cal O}(a^2)$ improved gauge action to examine the nature of the RW-endpoint 
at smaller than physical values of the light quark masses and a physical value 
of the strange quark mass.

\section{Lattice setup and simulation parameters}
We have performed our calculations with the HISQ action with a temporal extent 
of $N_\tau=4$ and with finite imaginary chemical potential corresponding to 
the value, $\mu/T=\pi/3$ for all quark flavors. The constant line of physics 
is defined according to \cite{Bazavov:2011nk}. It has been obtained by tuning 
the strange quark mass ($m_s$) for various values of the gauge couplings 
in such a way that we obtain the mass of the fictitious pseudo-scalar 
$\eta_{\bar{s}s}$ meson as $686$ MeV, which is same as obtained from the 
Gell-Mann-Okubo
mass formula in first order chiral perturbation theory. After tuning the 
strange quark mass $m_s$, the light quark masses $m_l$ is chosen such  that 
the $m_l/m_s$ ratio remains constant for all values of gauge couplings. Different 
ratios then correspond to different pion masses, e.g. $m_l/m_s=1/27$ 
corresponds to physical values of the quark masses and gives a 
pseudo-Goldstone pion mass of about $135$ MeV. The lowest quark mass ratio 
used in our calculations, $m_l/m_s=1/320$, corresponds to Goldstone pion mass 
of about $40$ MeV.
The partition function for $(2+1)$-flavor QCD with two degenerate light quark 
masses, $m_l=(m_u+m_d)/2$, a strange quark mass ($m_s$)
and identical chemical potentials $\mu/T=\pi/3$ for all flavors
may be written as,
\begin{equation}
	Z(T,V, \mu)=\int \mathcal{D}U\ {\rm det}[M_{l}(i\mu)]^{1/2}{\rm det}[M_s(i\mu)]^{1/4}\exp[-S_G] \; ,
\end{equation}
where, $M_f=D_{HISQ}(i\mu)+m_f$ is the fermion matrix for flavor $f$ and we
also use the notation $f\equiv l$ to denote one of the two degenerate
light quark flavors. For the gauge action, $S_G$, we use the tree 
level improved Symanzik action.
 
We use the kaon decay constant $f_k$ to set the temperature scale in 
MeV~\cite{Bazavov:2011nk}. We change the values of the gauge coupling for every 
ratio $m_l/m_s$ in such a way that the corresponding temperature varies in 
the range, $T\sim T_c\pm 0.1T_c$. Some further details on our simulation parameters 
are given in Table~\ref{tab:runs}.
\begin{table}[hbt]
	\centering
	\begin{tabular}{|c|c|c|c|}
		\hline
		$N_\sigma$ & $N_\tau$ & $m_l/m_s$ & $m_{\pi}$(MeV)\\
		\hline
		8 & 4 & 1/27 &  135\\ 
		\hline
		12 & 4 & 1/27 &  135\\
		\hline
		16 & 4 & 1/27, 1/40, 1/60 &  135, 110, 90 \\ 
		\hline
		24 & 4 & 1/27, 1/40, 1/160,1/320 &  135, 110, 55, 40 \\
		\hline
        32 & 4 & 1/160 &  55 \\
        \hline	
\end{tabular}
	\caption{Details of the numerical simulation parameters for calculations in $(2+1)$-flavor QCD on lattices of size $N_\sigma^3\times N_\tau$.}
	\label{tab:runs}
\end{table}


\section{Results}
\subsection{Nature of the RW endpoint}
All our calculations have been performed in the RW plane, {\it i.e.} in the 
absence of a symmetry breaking field $h=\mu-\mu_{RW}$, that would break 
explicitly the $Z(2)$ symmetry of the QCD Lagrangian in the RW plane. 
The appropriate
approach for studying the RW phase transition in this case is using
finite size scaling relations. As the expectation value of the imaginary
part of the Polyakov loop vanishes for $h=0$, one usually uses its 
absolute value as an order parameter,
\begin{equation}
        M\equiv \langle |{\rm Im}~P|\rangle =
	\left\langle \left|\frac{1}{N_\sigma^3} \sum_{\bf x}{\rm Im}~
	{\rm Tr} \prod_{x_0=1}^{N_\tau} U_{x,\hat{0}} \right| \right\rangle
        \, .
        \label{absPolyakov}
\end{equation}
The finite size scaling relations for the order parameter $M$ and 
its susceptibility, 
\begin{equation}
\chi_M = N_\sigma^3 \left( 
\langle ({\rm Im}~P)^2\rangle  -\langle |{\rm Im}~P|\rangle^2 \right)\, ,
        \label{susM}
\end{equation}
are expressed in terms of two universal finite-size scaling functions, 
$f_{G,L}$ and $f_{\chi,L}$, of
the $3$-$d$, $Z(2)$ universality class,
\begin{eqnarray}
M &=& A N_\sigma^{-\beta/\nu} f_{G,L} (z_f) + \; reg. \; ,
\label{M_scaling}
\\
\chi_M &=& B N_\sigma^{\gamma/\nu}~f_{\chi,L}(z_f) + \; reg. \; ,
\label{sus_scaling}
\end{eqnarray}
where $z_f=z_0 tN_\sigma^{1/\nu}$ and $t=(T-T_c)/T_c$.
These are related to the  infinite volume 
universal scaling functions of the order parameter and its susceptibility, 
$f_G$ and $f_\chi$, respectively \cite{Engels:2002fi}. 
We also calculate the Binder cumulant \cite{Binder:1981sa,Binder:1981zz} $B_4$ 
which is given in terms of the ratio of fourth and second order moments of
the order parameter,
\begin{eqnarray}
B_4(T,V)&=&\dfrac{\langle(Im~P)^4\rangle}{\langle (Im~P)^2\rangle^2}\; .
\end{eqnarray}
Close to the critical point
this ratio also is expressed in terms of a universal ratio of scaling
functions, $B_4(T,V)= f_B(z_f) +\ {\rm reg.}$.

For our finite-size
scaling analysis we have calculated the finite-size scaling functions
using the $\lambda \phi^4$ spin model with a tuned parameter $\lambda$
such as to reduce contributions arising from corrections-to-scaling 
(improved Ising model). This $Z(2)$ invariant spin model has been used
previously to determine the infinite volume scaling functions of the $3$-$d$
Ising model \cite{Engels:2002fi}.

In order to test universality of several observables obtained from our
$(2+1)$-flavor QCD calculations with an imaginary chemical potential
we only fit the order parameter with the universal scaling function $f_{G,L}$.
This fit yields results for the non-universal parameters 
$T_c, A ~\text{and}~z_0$. We use these parameters to compare other observables,
e.g. $\chi_M$ and $B_4$ to the corresponding scaling functions $f_\chi$, $f_B$.
These comparisons are parameter free, when taking into account that the 
amplitude $B$, appearing in Eq.~\ref{sus_scaling}, can be expressed in terms
of the amplitude $A$ in Eq.~\ref{M_scaling}, {\it i.e.} $B=A^2$.

\begin{figure}[t]
	\hspace*{-0.33cm}\includegraphics[scale=0.43]{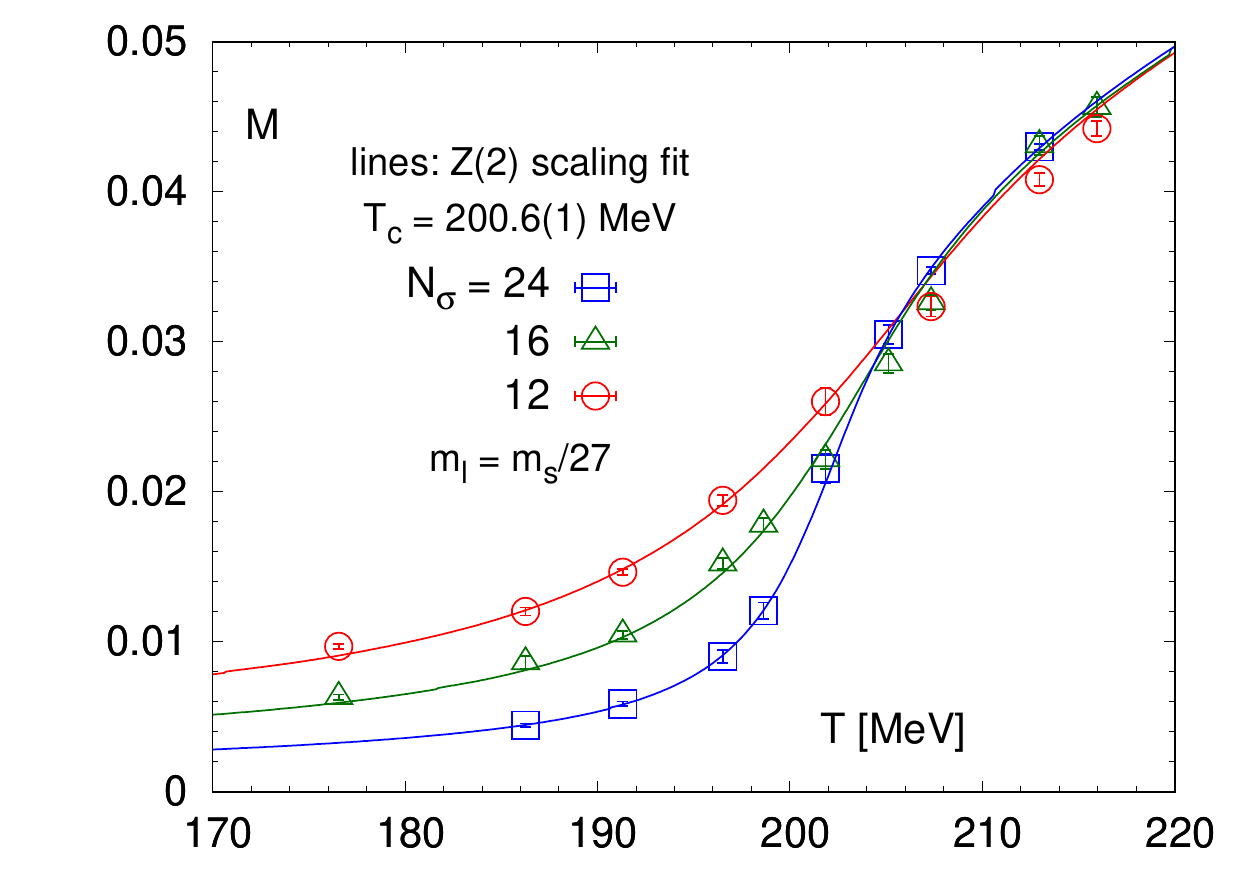}\hspace*{-0.41cm}
        \includegraphics[scale=0.43]{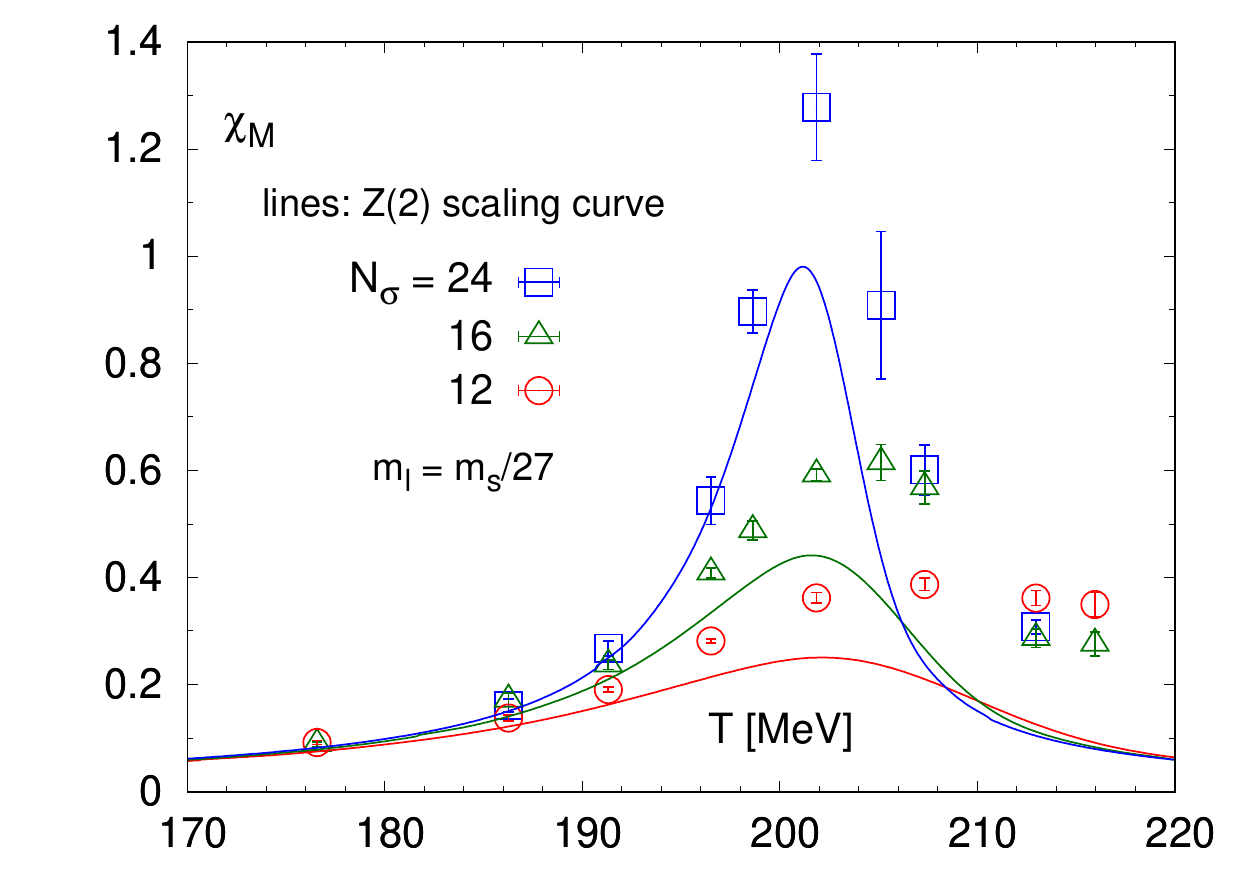}\hspace*{-0.41cm}
	\includegraphics[scale=0.43]{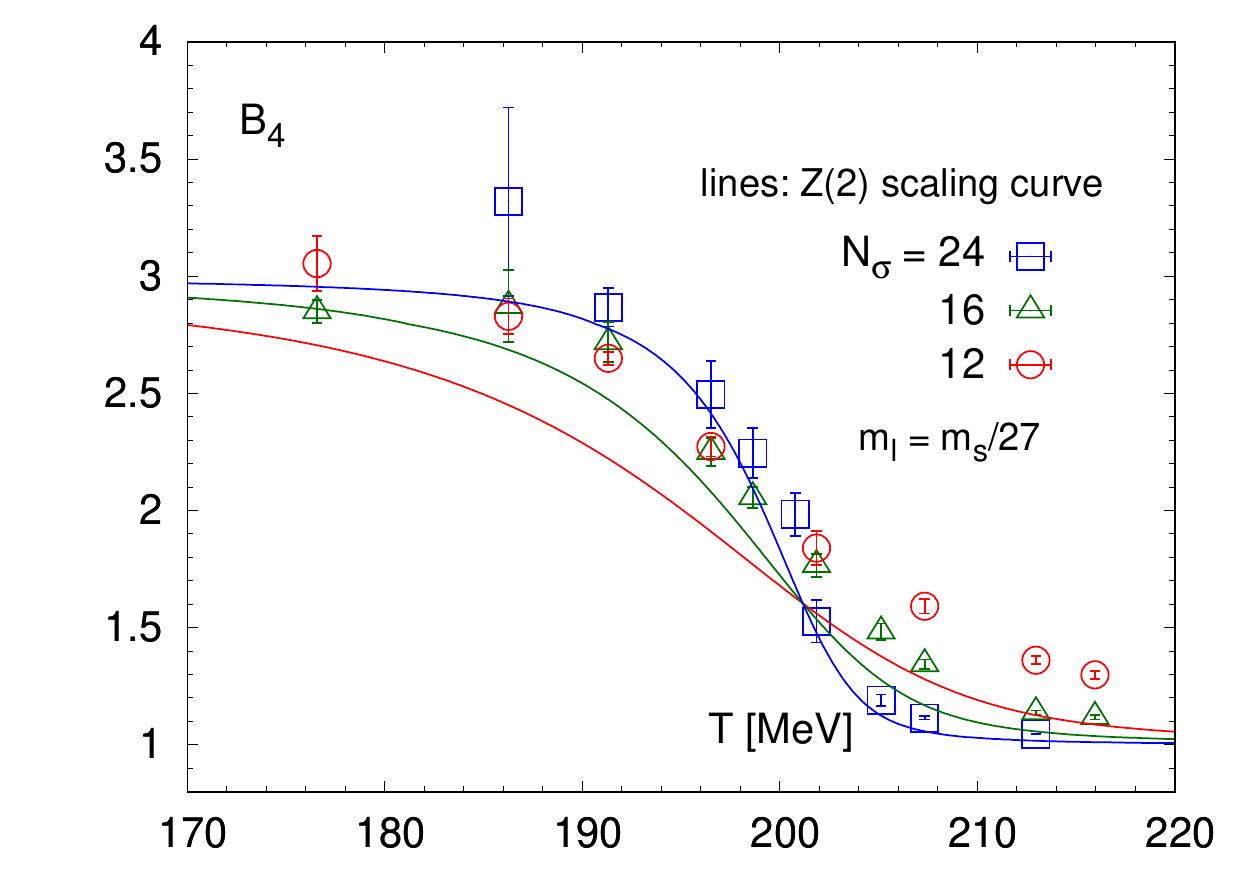}
        \caption{The order parameter $M$ (left), its susceptibility $\chi_M$ 
	(middle) and the Binder cumulant $B_4$ (right) for $m_l/m_s=1/27$ and 
various lattice sizes. Lines are based on a fit of $M$ to the $Z(2)$ 
finite-size scaling function $f_{G,L}$. Curves shown for $\chi_M$ and $B_4$
are then parameter free results of this fit.}
   \label{fig:op_sus}
\end{figure}

\begin{figure}[t]
        \centering
       \hspace*{-0.3cm}\includegraphics[scale=0.42]{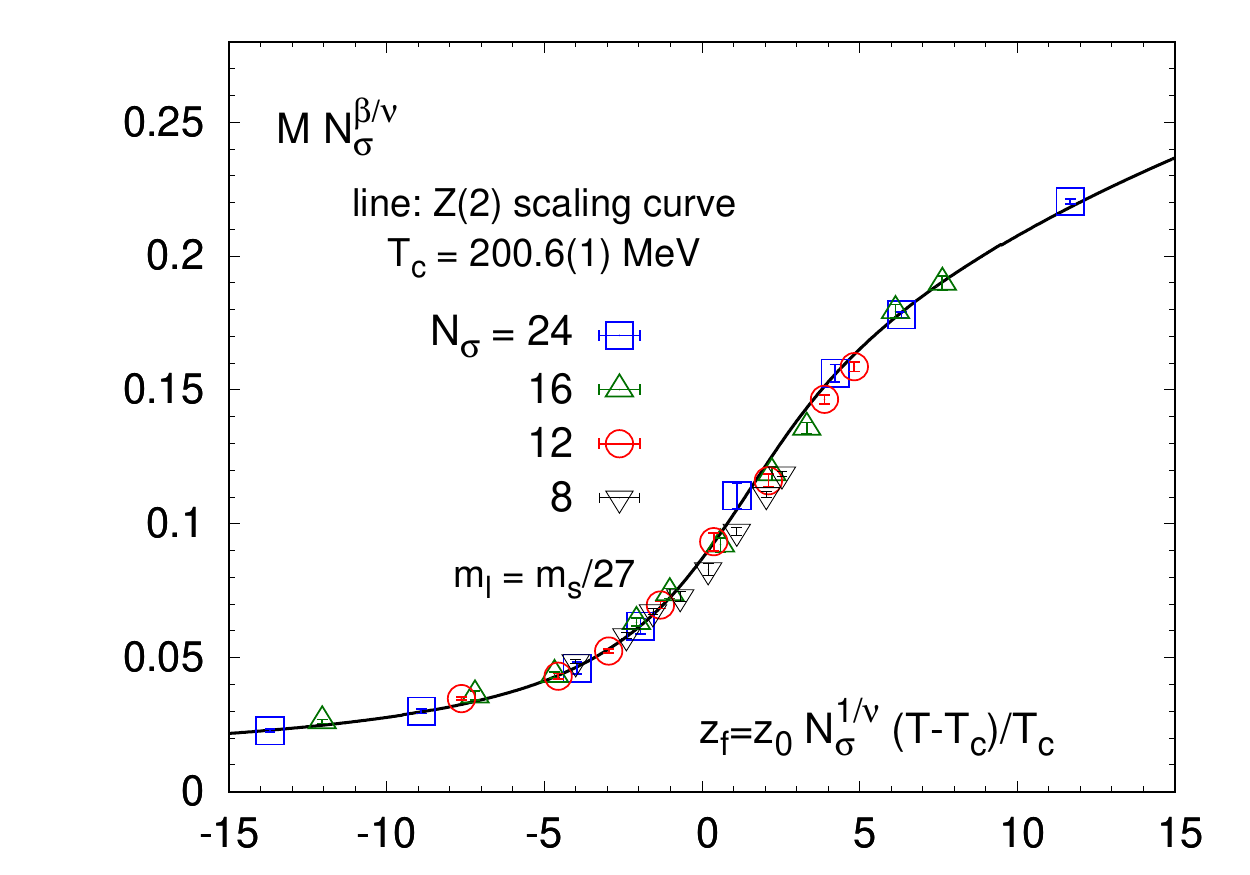}\hspace*{-0.25cm}
        \includegraphics[scale=0.42]{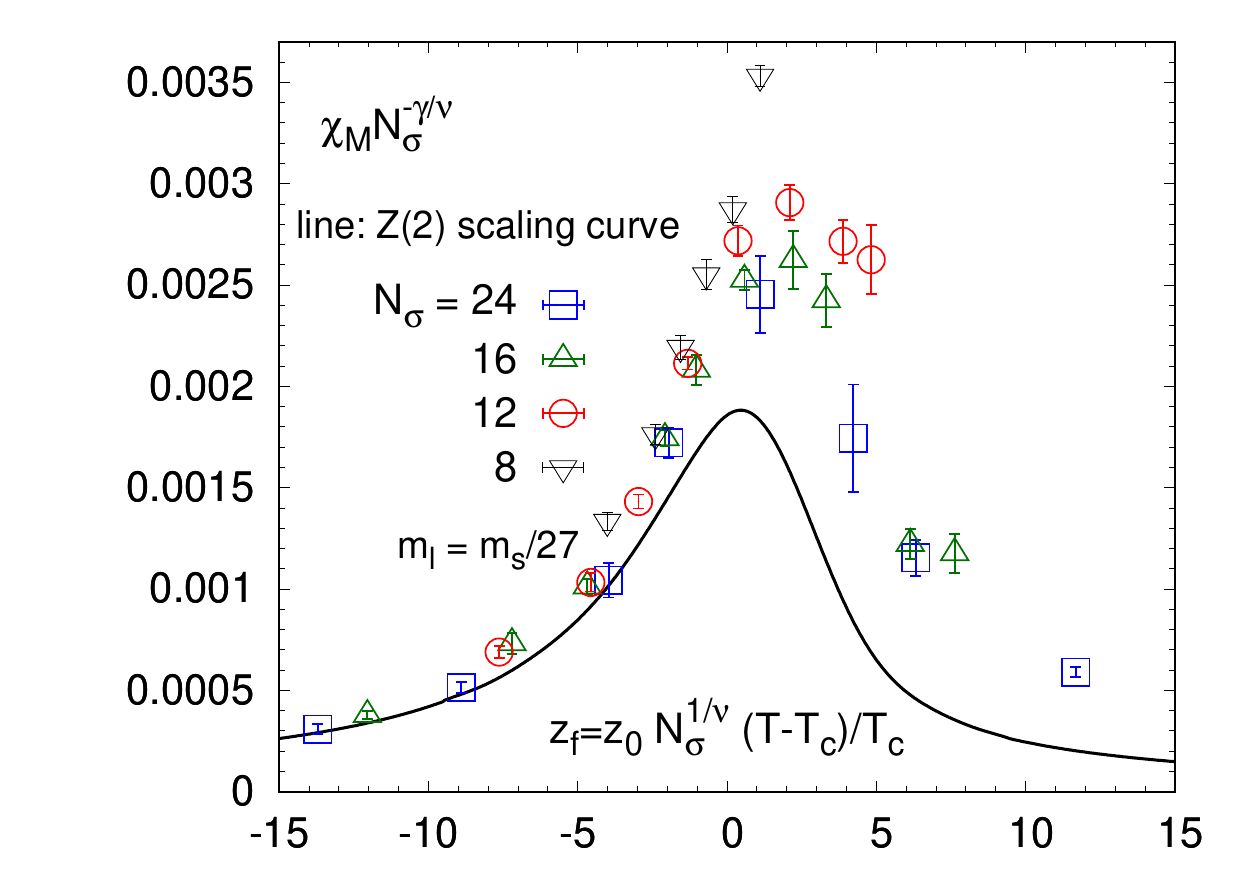}\hspace*{-0.25cm}
        \includegraphics[scale=0.42]{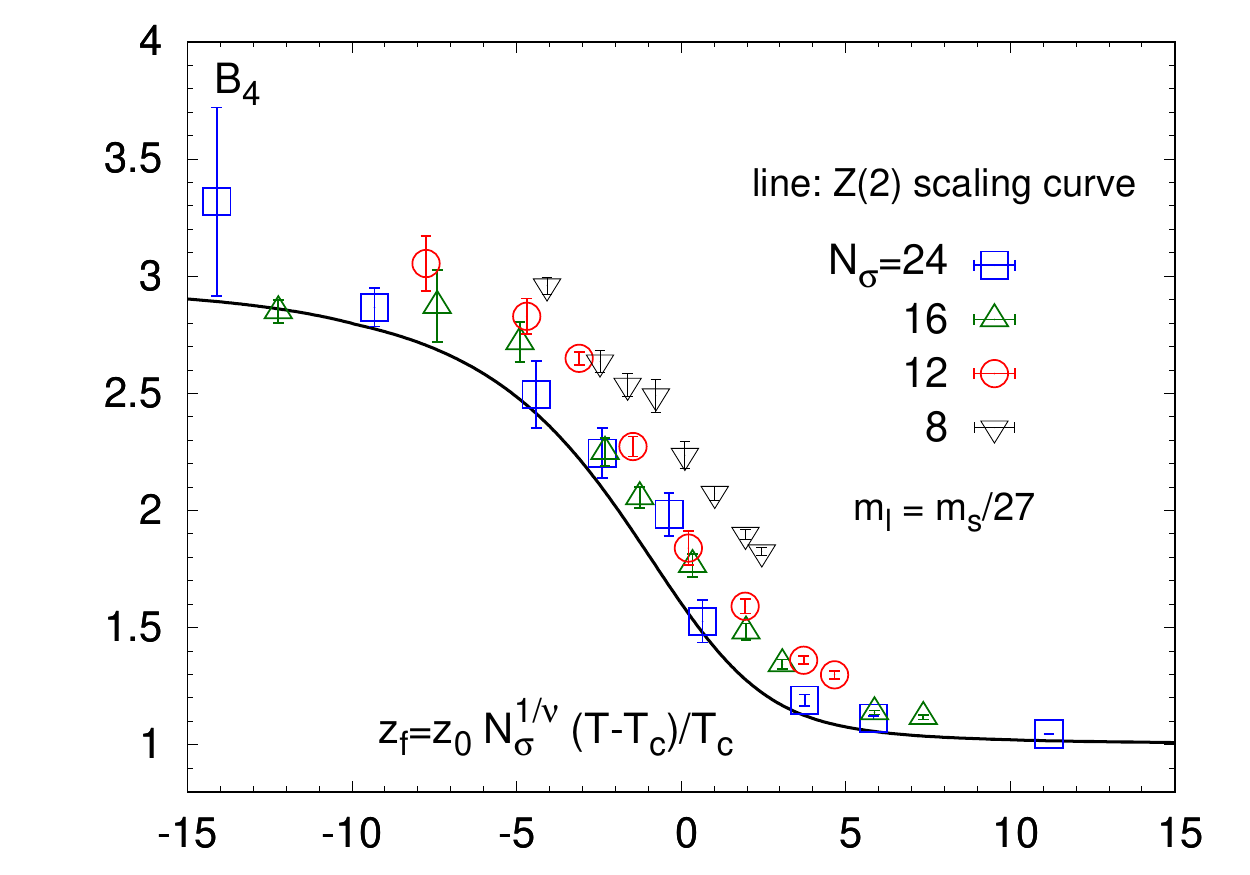}
        \vspace{-0.2cm}
        \caption{The rescaled order parameter $M$ (left), the rescaled 
		susceptibility $\chi_M$
        (middle) and the rescaled Binder cumulant $B_4$ (right) 
	for $m_l/m_s=1/27$ and various lattice sizes. Lines show the 
corresponding $3$-$d$, $Z(2)$ finite size scaling functions.}
        \label{fig:binder_k}
\end{figure}

In Fig.~\ref{fig:op_sus} we show the order parameter $M$, its 
susceptibility $\chi_M$ and the Binder cumulant $B_4$, calculated on lattices
with different spatial extent, $N_\sigma$, and for the physical light to 
strange quark mass ratio $m_l/m_s=1/27$. Also shown are fits to $M$ based on 
the ansatz given in Eq.~\ref{M_scaling} and the resulting finite-size 
scaling curves for $\chi_M$.
The order parameter data follow the universal scaling curve nicely in the 
explored temperature range. For the susceptibility, which is not fitted,
scaling  violations are clearly visible in the symmetry broken phase for the 
smaller volumes, while scaling violations seem to be small in the symmetric
phase. A similar behavior also is found for the Binder cumulant $B_4$. 
The significance of scaling violations on small lattices also is 
apparent 
from the rescaled observables shown in Fig.~\ref{fig:binder_k}. 

When changing the spatial extent, $N_\sigma$, of the lattice, the peak height 
of the order parameter susceptibility
would increase proportional to volume, $V\equiv N_\sigma^3$, if the phase
transition at the RW endpoint would be first order in the infinite volume 
limit. For a second order phase transition in the $3$-$d$, $Z(2)$ universality 
class the peak height 
only increases as $N_\sigma^{\gamma / \nu}= N_\sigma^{1.96}$. As can be 
seen from the rescaled susceptibility, Fig.~\ref{fig:binder_k}~(middle), 
our results for $m_l/m_s=1/27$ are
consistent with this slower rise. We thus find that the RW endpoint
for this quark mass ratio still corresponds to a $2^{nd}$ order phase
transition in the Ising universality class. 

\begin{table}[b]
        \centering
        \begin{tabular}{|c|c|c|c|}
                \hline
                $m_l/m_s$ & $m_\pi$ & $T_c$ & $\Delta{T_c}$\\
                \hline
                 1/27 & 135 &200.6 & 0.18 \\
                \hline
                 1/40 & 110 & 199.37 & 0.23 \\
                \hline
             1/160 & 55 & 195.23 & 0.27 \\
                \hline
                1/320 & 40 &195.36 & 0.29 \\
            \hline
\end{tabular}
        \caption{Phase transition temperature, $T_c$, at the endpoint
		of the line of $1^{st}$ order transitions in the RW plan.
	Results are obtained on lattices with temporal extent $N_\tau=4$ an
        different quark mass ratios $m_l/m_s$.}
        \label{tab:results}
\end{table}

Similar behavior we find in all our calculations with light quark masses
down to $m_l=m_s/320$. The non-universal scaling parameters, of course,
will be quark mass dependent. Indeed, we find that the phase transition
temperature decreases slightly with quark mass, although we cannot yet
specify in detail the functional form of the dependence on $m_l/m_s$.
Results for the phase transition temperature at different values of
$m_l/m_s$ are summarized in Table~\ref{tab:results}.

\begin{figure}[t]
	\centering
	\includegraphics[scale=0.42]{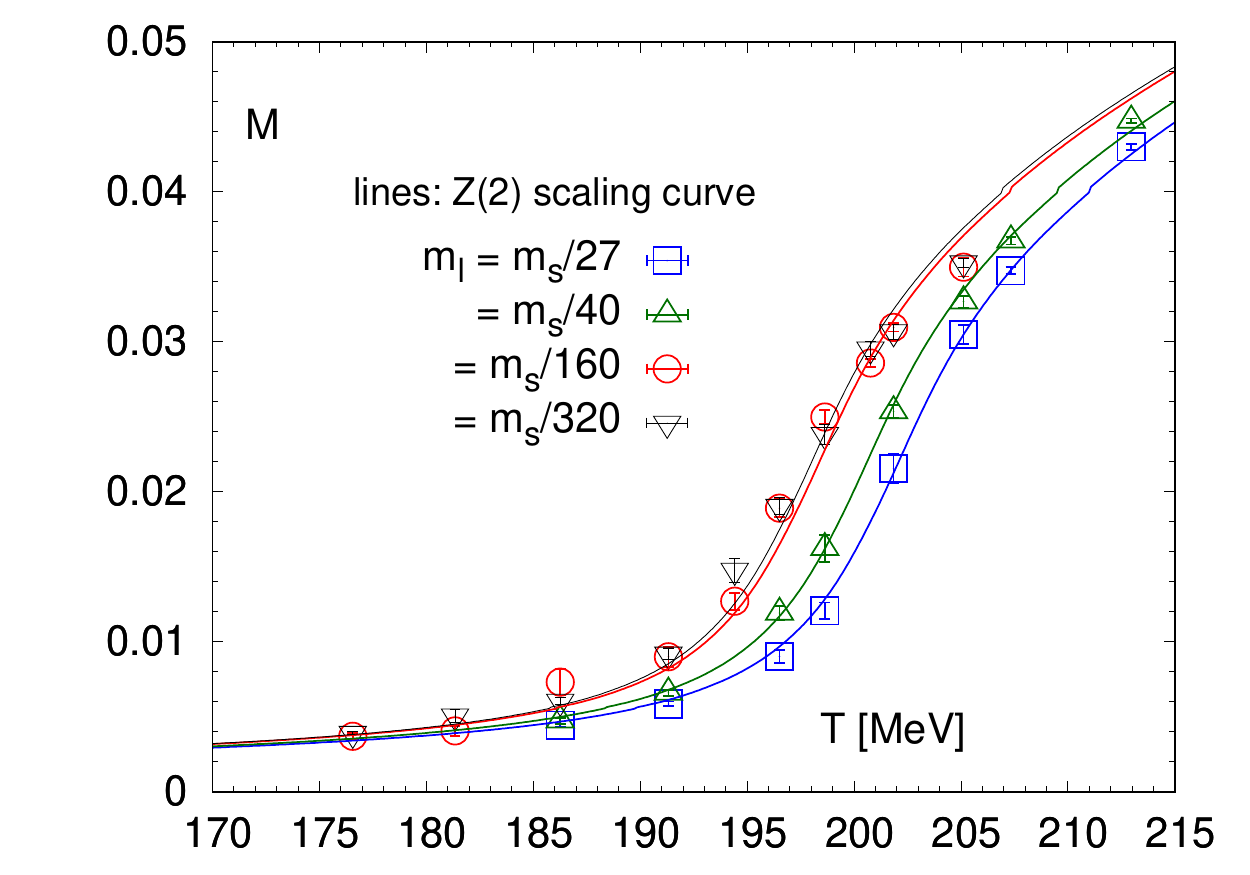}
	\includegraphics[scale=0.42]{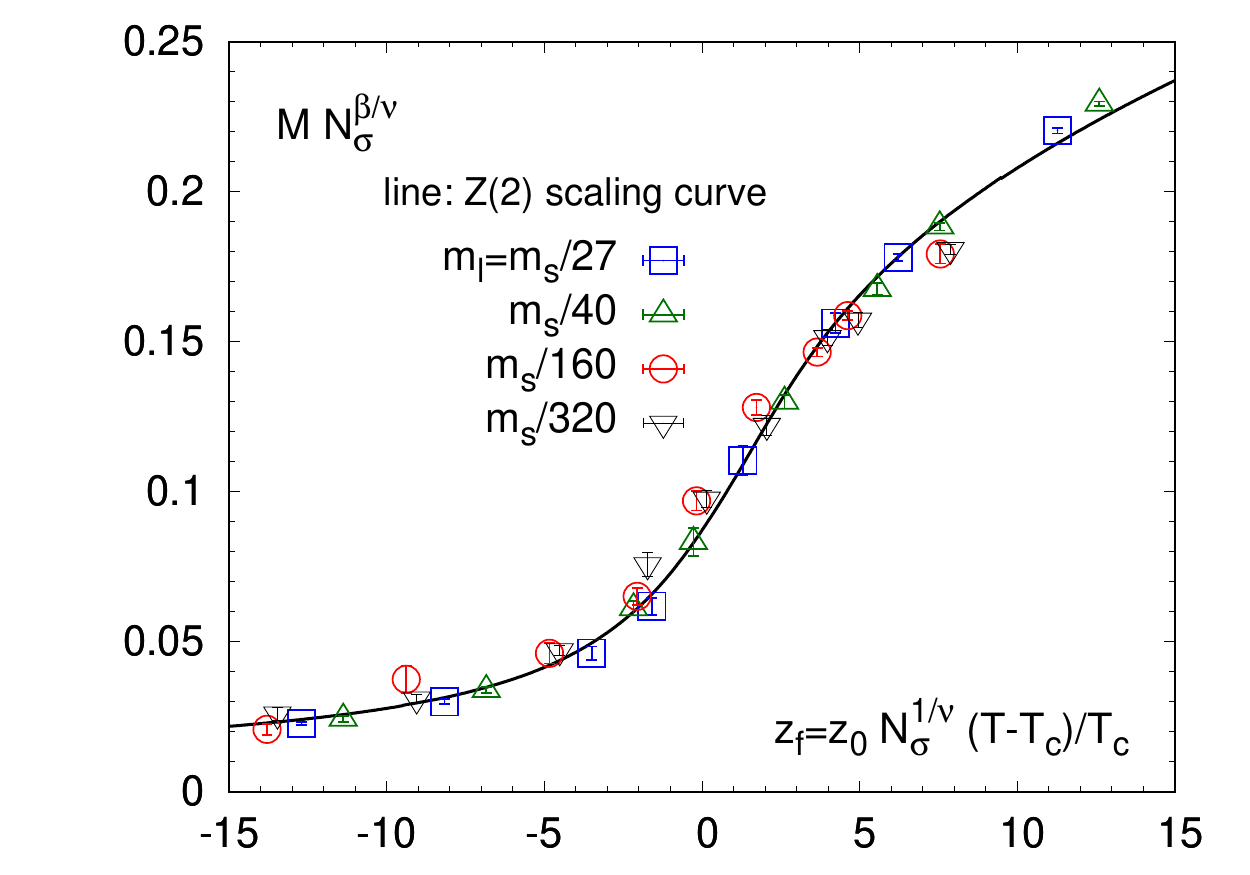}
\caption{The order parameter $M$ for various $m_l/m_s$ ratios for a fixed volume $N_\sigma=24$. Lines correspond to fits using the Z(2) finite size scaling functions. The corresponding rescaled data are shown in the figure on the right.}
\label{fig:op_chiral}
\end{figure}

Results obtained for the order parameter $M$ on lattices with spatial
extent $N_\sigma=24$ and for four different values of the light quark
masses are shown in Fig.~\ref{fig:op_chiral}~(left). The shift of the
crossover region to smaller values of the temperature is obvious, while the
shape of the overall temperature dependence does not show any significant 
modification, indicating that any quark mass dependence of the non-universal
amplitude $A$ and the scale factor $z_0$ is small. In fact, we do not
find any systematics for the quark mass dependence of these parameters.
As can be seen in Fig.~\ref{fig:op_chiral}~(right) the rescaled order
parameter in all cases agrees well with the universal $3$-$d$, $Z(2)$
scaling function $f_{G,L}(z_f)$, giving further support to the
assertion that the RW phase transition is $2^{nd}$ order in the entire
interval of quark masses considered by us. 

\subsection{Chiral phase transition in RW plane}
\begin{figure}
        \centering
        \hspace*{-0.33cm}\includegraphics[scale=0.42]{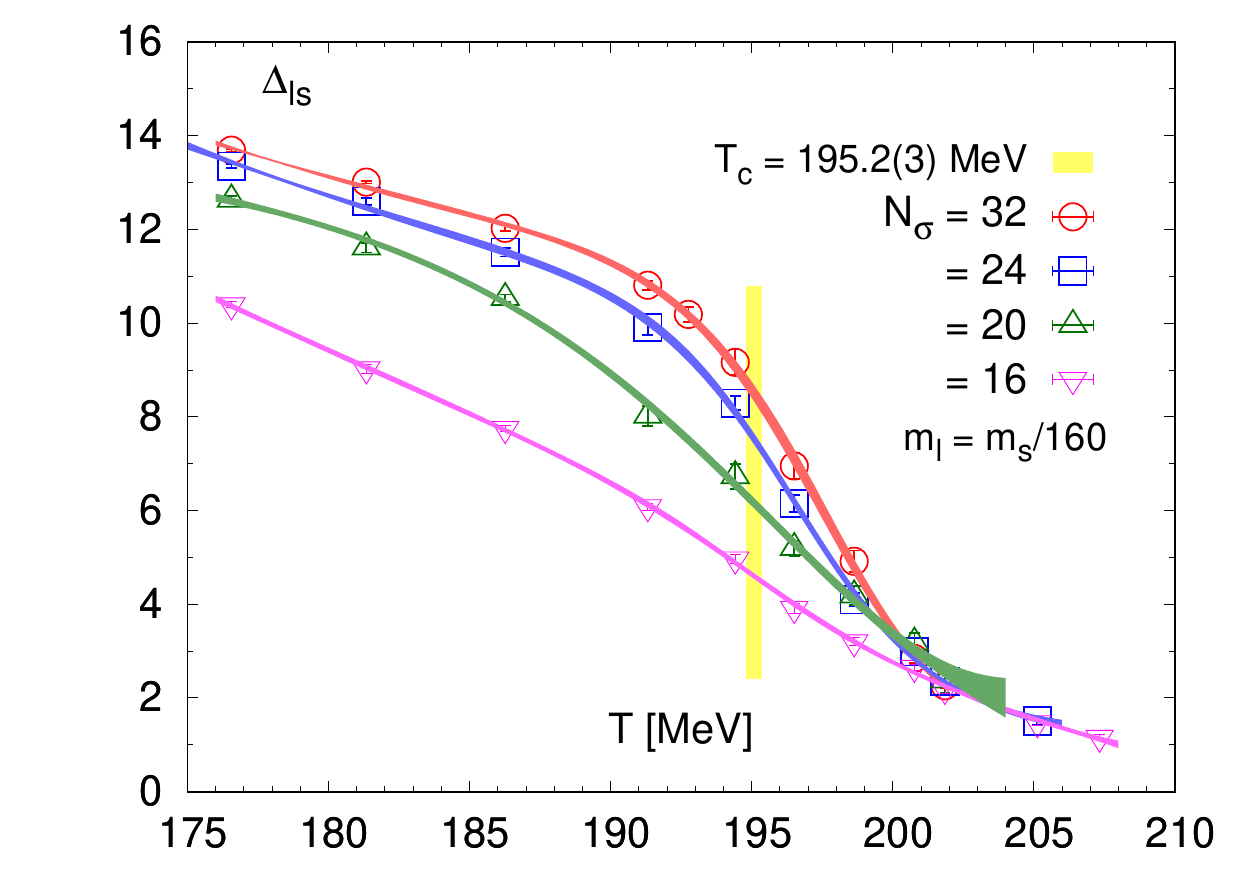}\hspace*{-0.41cm}
        \includegraphics[scale=0.42]{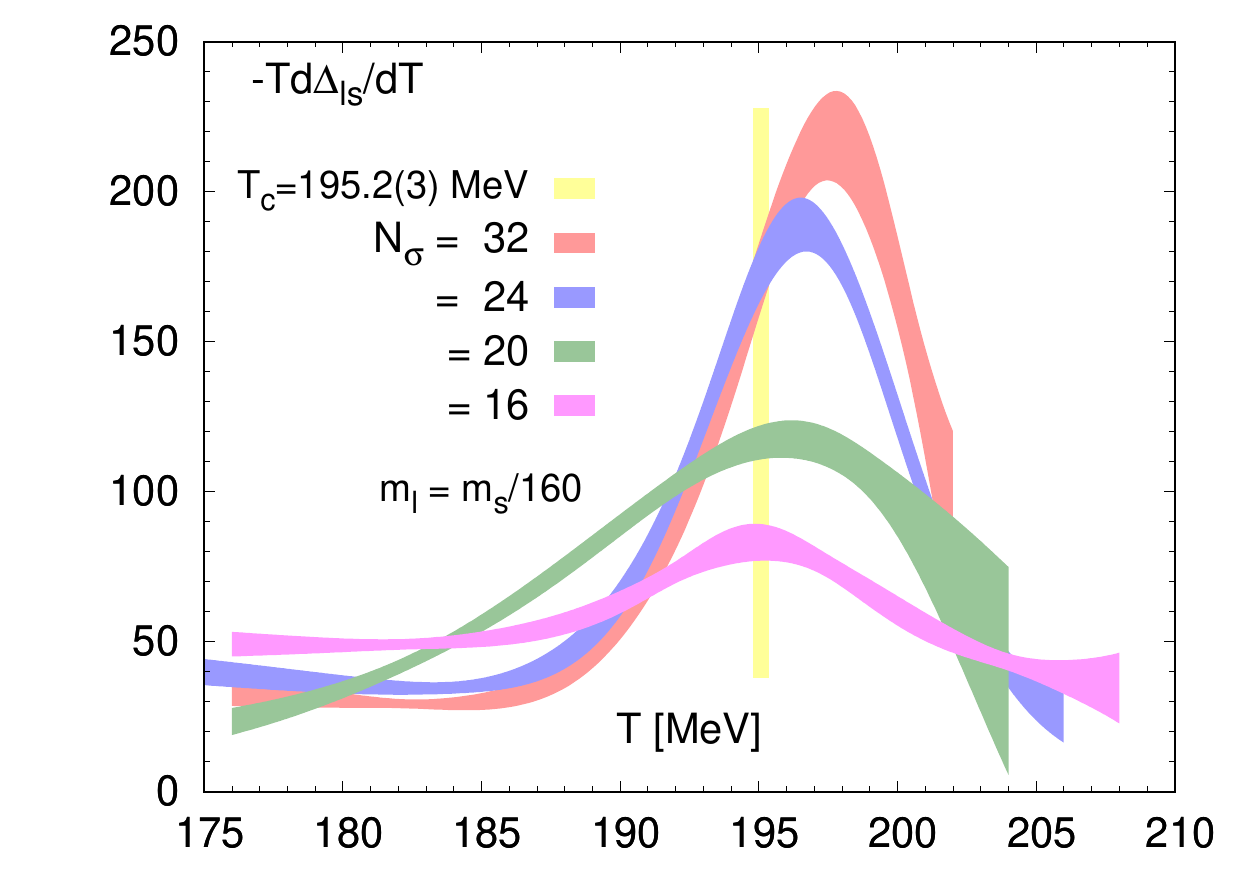}\hspace*{-0.41cm}
        \includegraphics[scale=0.42]{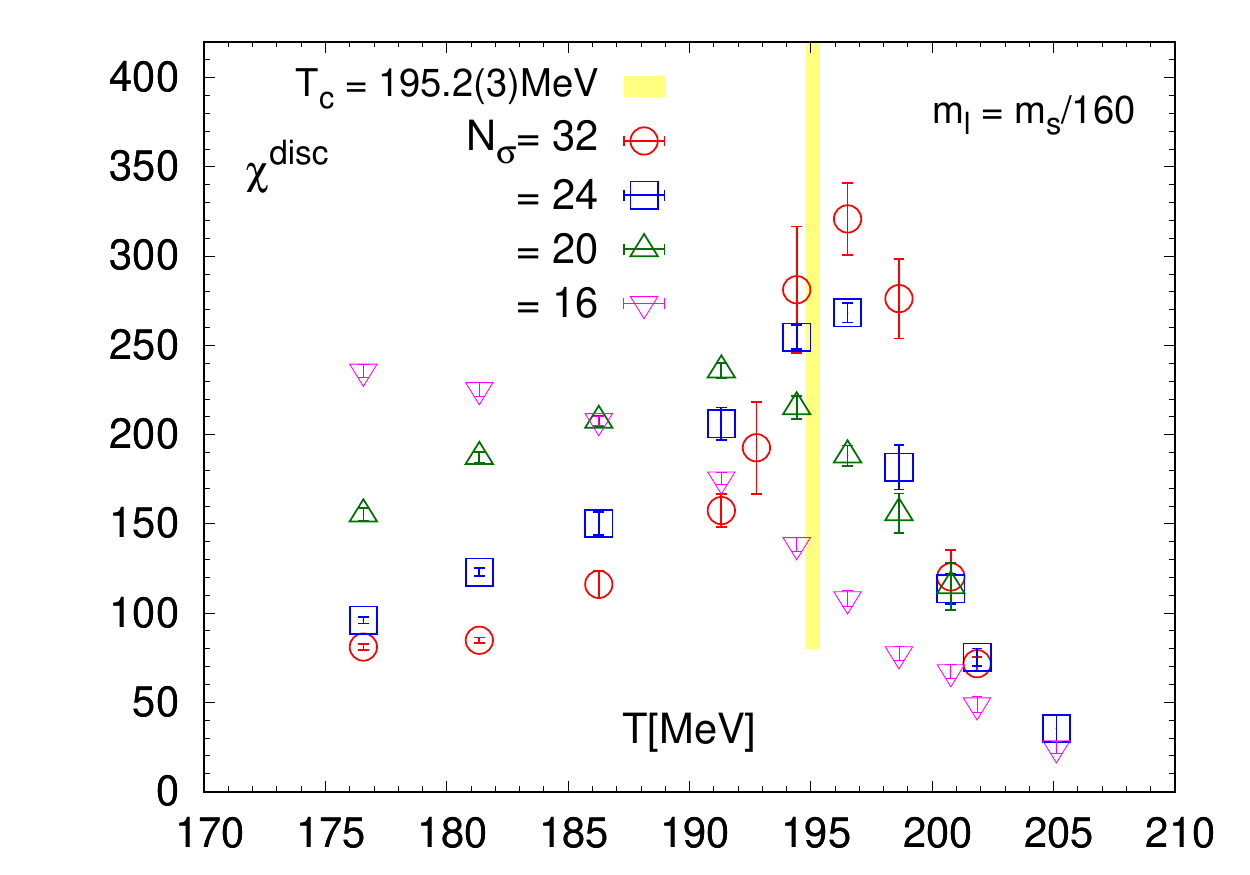}
	\caption{The RGI chiral condensate, $\Delta_{ls}$ (left), 
	its temperature derivative (middle), 
	and the disconnected chiral susceptibility (right)
	versus temperature for various spatial lattice sizes and fixed quark 
	mass ratio  $m_l/m_s=1/160$. The yellow bands show the location of
	the RW transition temperature extracted from the location of the peak 
	of the order parameter susceptibility $\chi_M$.
	         }
\label{fig:Rw_chiral1}
\end{figure}

Let us now turn to a discussion of the chiral phase transition in $(2+1)$-flavor
QCD at non-zero imaginary chemical potential.
In the chiral limit the chiral phase transition occurs alongside with the RW 
transition. The order parameter for this transition, the renormalization group 
invariant (RGI) chiral condensate, can be defined as,
 \begin{equation}
\Delta_{ls} = \frac{2}{f_K^4} \left( m_s \langle \bar{\psi}\psi \rangle_l - 
m_l \langle \bar{\psi}\psi \rangle_s\right) \; ,
\end{equation}
where 
\begin{eqnarray}
\langle \bar{\psi}\psi \rangle_f =
\frac{T}{V}~\frac{\partial \ln\  Z(T,V,m_f,\mu)}{\partial m_f} =
\frac{1}{N_\sigma^3N_\tau}\langle \rm{Tr} M_f^{-1}\rangle\; .
\end{eqnarray}
We used the kaon decay constant $f_K$ for normalization of the dimensionless
order parameter $\Delta_{ls}$.
For the determination of a pseudo-critical temperature, $T_\chi$, for chiral 
symmetry restoration we use the fluctuations of the light quark chiral 
condensate,
\begin{eqnarray}
\chi^{disc}=4\frac{m_s^2}{N_\sigma^3N_\tau}\left( 
\langle({\rm{Tr} M_l^{-1})^2\rangle-\langle( Tr M_l^{-1})\rangle^2}\right)/f_k^4
\; .
\end{eqnarray}
In Fig.~\ref{fig:Rw_chiral1}  we show the order parameter
$\Delta_{ls}$, its temperature derivative $Td\Delta_{ls}/dT$, 
and the disconnected chiral susceptibility $\chi^{disc}$ 
for the light quark mass, $m_l=m_s/160$ . 
The weak volume dependence found for the peak height of $Td\Delta_{ls}/dT$
as well as $\chi^{disc}$ rules 
out the possibility of a first order RW transition in the chiral
observables. This is consistent with the conclusion from the previous section. 
Nonetheless, the peak heights continue to increase with increasing volume
also on the largest lattices available to us.  
This suggests that the chiral susceptibilities at non-zero values
of the quark mass, are sensitive to the RW transition. In fact, they are
expected to behave like temperature like observables in an effective
$Z(2)$ symmetric Hamiltonian for the RW transition. The slow rise of
the peak heights is consistent with this assertion.
Also the location of the peak of $\chi^{disc}$, 
$T_\chi$ is consistent with the phase transition temperature, $T_c$, of
the RW transition. This suggests that also in the chiral limit 
the RW and chiral transition may coincide and occur at the same temperature. 
However, a more detailed analysis of such a scenario is clearly needed.

\section{Conclusions}
\vspace{-0.2cm}
We presented here some results from our ongoing studies of the phase transition
in $(2+1)$-flavor QCD with imaginary chemical potential in the RW plane. Our 
finite size scaling analysis with the HISQ action, performed on lattices with
temporal extent $N_\tau=4$, show that the phase transition in the RW plane
is $2^{nd}$ order belonging to the $3$-$d$, $Z(2)$ universality class for
physical values of the quark masses down to small quark masses, corresponding 
to a Goldstone pion mass of about $40$~MeV.
At present we see no indication for a possible $1^{st}$ order transition
that may occur in the chiral limit. However, the nature of the chiral 
phase transition itself still requires further analysis. Current results 
indicate that it may coincide with the RW transition also in the chiral limit.

\section*{Acknowledgement}
\vspace{-0.2cm}
This work was supported in part through Contract No. DE-SC001270 with the
U.S. Department of Energy,
the Deutsche Forschungsgemeinschaft (DFG) through the CRC-TR 211
"Strong-interaction matter under extreme conditions", grant number
315477589 - TRR 211,
and the grant 05P18PBCA1 of the German Bundesministerium f\"ur Bildung und
Forschung.
\vspace{-0.2cm}
\bibliographystyle{JHEP}
\bibliography{cpod2018}
\end{document}